\documentclass[a4paper,11pt]{article}
\usepackage{jheppub} 
\usepackage{lineno}
\usepackage{cancel}
\usepackage{xcolor}
\usepackage{abraces}
\usepackage{amsfonts,amsmath, amssymb, mathtools}
\usepackage{stackengine}
\usepackage{mathrsfs}
\usepackage{commath}
\usepackage[utf8]{inputenc}
\usepackage{fourier} 
\usepackage{array}
\usepackage{makecell}
\newcommand{\NN}{\mathcal{N}}

\usepackage{tikz}
\usetikzlibrary{shapes.geometric, arrows}

\tikzstyle{startstop} = [rectangle, rounded corners, 
minimum width=3cm, 
minimum height=1cm,
text centered, 
draw=black, 
fill=red!30]

\tikzstyle{io} = [trapezium, 
trapezium stretches=true, 
trapezium left angle=70, 
trapezium right angle=110, 
minimum width=3cm, 
minimum height=1cm, text centered, 
draw=black, fill=blue!30]

\tikzstyle{process} = [rectangle, 
minimum width=3cm, 
minimum height=1cm, 
text centered, 
text width=3cm, 
draw=black, 
fill=orange!30]

\tikzstyle{decision} = [diamond, 
minimum width=3cm, 
minimum height=1cm, 
text centered, 
draw=black, 
fill=green!30]
\tikzstyle{arrow} = [thick,->,>=stealth]

\title{\boldmath Scalar-Tensor multiplet in four dimensional $\mathcal{N} =2$  conformal supergravity}

\author{Aravind Aikot and Bindusar Sahoo}
\affiliation{School of Physics, Indian Institute of Science Education and Research Thiruvananthapuram,\\
	Vithura, Thiruvananthapuram, India}
\emailAdd{arvd1719@alumni.iisertvm.ac.in, bsahoo@iisertvm.ac.in}
\abstract{We study various $\NN=2$ multiplets in four dimensions by looking at the supersymmetric truncation of four dimensional $\NN=3$ multiplets. Under supersymmetric truncation, the off-shell $\mathcal{N}=3$ Weyl multiplet reduces to the off-shell $\mathcal{N}=2$ Weyl multiplet and the off-shell $\mathcal{N}=2$ vector multiplet (which we will refer to as the central charge multiplet). Under the same truncation, the on-shell $\mathcal{N} =3$ vector multiplet reduces to the on-shell $\mathcal{N} = 2$ vector multiplet and an on-shell massive hypermultiplet with a broken rigid $SU(2)$ and a non-trivial central charge transformation. We use the field equations of this hypermultiplet to eliminate some of the fields of the central charge multiplet in terms of the fields of the hypermultiplet and a dual tensor gauge field (similar in spirit to how a dilaton Weyl multiplet is constructed). This results in a new off-shell matter multiplet, with $8+8$ degrees of freedom, containing scalar fields  and a tensor gauge field, which we refer to as the scalar-tensor multiplet.} 

\begin{document}
\maketitle
\flushbottom
\vspace{-1cm}

\section{Introduction}
Conformal supergravity is an important technical framework which facilitates the construction of Poincar{\'e} supergravity theories with or without higher derivative corrections in a systematic and tractable manner. The whole procedure of constructing Poincar{\'e} supergravity theories from conformal supergravity theories is greatly facilitated by the use of various multiplets. The multiplet that plays a very crucial role in this construction is the Weyl multiplet which contains the graviton, its superpartner gravitino as well as other gauge fields corresponding to the underlying superconformal algebra. It is an off-shell multiplet and apart from the gauge fields, it also contains several auxiliary matter fields for the off-shell closure of the multiplet. Apart from the Weyl multiplet, there are several matter multiplets which also play a crucial role. The most important role played by these matter multiplets is that of a compensating multiplet that allows one to go from conformal supergravity to Poincar{\'e} supergravity. Since Poincar{\'e} supergravity has less symmetries than conformal supergravity, the degrees of freedom of a Poincar{\'e} supergravity multiplet is understandably more than the Weyl multiplet. The additional degrees of freedom are provided by the compensating matter multiplets. For $\NN=3$ as well as $\NN=4$ supergravity in four dimensions, the on-shell vector multiplet plays the role of the compensating multiplets \cite{Hegde:2022wnb,deRoo:1984zyh}. For $\NN=2$ supergravity in four dimensions, two compensating multiplets are required. One of them is always a vector multiplet which provides the graviphoton whereas the second compensating multiplet can be either a linear, non-linear or a hypermultiplet leading to different formulations of $\NN=2$ supergravity in four dimensions \cite{deWit:1980lyi}.\footnote{The authors of \cite{Galperin:1987ek,Galperin:2001seg} have discussed the most general matter coupled $\NN=2$ Poincar{\'e} supergravity which is obtained by compensating the $\NN=2$ conformal supergravity with an off-shell hypermultiplet described in harmonic superspace with infinite number of auxiliary fields. This version is dubbed as the principal version and the authors have further shown that all other versions obtained using compensators with finite number of auxiliary fields can be obtained from the principal version by appropriate duality transformations.} $\NN=2$ conformal supergravity is very rich in multiplets, unlike its $\NN=3$ and $\NN=4$ cousins. Some of the multiplets that are well known in $\NN=2$ conformal supergravity are the vector multiplet \cite{deWit:1980lyi}, chiral multiplet \cite{deWit:1980lyi}, tensor or linear multiplet \cite{deWit:1980lyi}, non-linear multiplet \cite{deWit:1980lyi}, scalar or the hypermultiplet \cite{deWit:1980lyi}, real scalar multiplet \cite{Hegde:2017sgl} and the relaxed hypermultiplet \cite{Hegde:2020gpt}. 

Some of the above mentioned multiplets can be obtained from the supersymmetric truncation of $\NN=3$ multiplets or double supersymmetric truncation of $\NN=4$ multiplets. Supersymmetric truncation is a procedure where one rewrites the multiplets with higher supersymmetries in terms of multiplets with lower supersymmetries and consistently truncate some of the multiplets and any gauge transformation corresponding to such multiplets, thereby obtaining the desired multiplet with lower supersymmetry. For example, it has been shown that one can obtain $\NN=3$ Weyl multiplet from $\NN=4$ Weyl multiplet by consistently truncating the components which would have otherwise formed an $\NN=3$ gravitino multiplet \cite{vanMuiden:2017qsh,Hegde:2022wnb}.\footnote{The authors of \cite{vanMuiden:2017qsh} truncate the $\NN=4$ Weyl current multiplet to obtain the $\NN=3$ Weyl current multiplet and obtain the $\NN=3$ Weyl multiplet from the current multiplet.} A similar procedure has also been employed in \cite{Yamada:2019ttz} to obtain $\NN=1$ multiplets from $\NN=2$ multiplets.

In the first part of this work, we employ supersymmetric truncation procedure on the $\NN=3$ Weyl multiplet to obtain the off-shell $\NN=2$ Weyl multiplet and the off-shell $\NN=2$ vector multiplet. We will refer to this off-shell $\NN=2$ vector multiplet as the central charge multiplet and the corresponding $U(1)$ gauge transformation as the central charge transformation. We will also perform a supersymmetric truncation on the $\NN=3$ vector multiplet. This will give rise to two multiplets which can be identified with an on-shell $\NN=2$ vector multiplet and an $\NN=2$ hypermultiplet. However, this hypermultiplet comes with a central charge transformation which breaks the rigid $SU(2)$ symmetry of the hypermultiplet. 

In the second part of the work, we will use the hypermultiplet field equations coupled to the central charge multiplet to solve for some of the components of the central charge multipet in terms of the fields of the hypermultiplet and also a dual tensor gauge field. In this way, we will arrive at the $\NN=2$ scalar-tensor multiplet.

The plan of the paper is as follows. In section-\ref{N3sugra}, we will give a brief overview of $\NN=3$ conformal supergravity and the $\NN=3$ multiplets. In section-\ref{susytrunc}, we will perform supersymmetric truncation of the $\NN=3$ multiplets. In section-\ref{N2hyper}, we will discuss about an off-shell $\NN=2$ hypermultiplet which has a rigid $SU(2)$ symmetry and show that upon a specific choice of central charge transformation, we get one of the multiplets obtained in the previous section via supersymmetric truncation of an $\NN=3$ vector multiplet. We will also show that this choice of central charge transformation breaks the rigid $SU(2)$ of the off-shell hypermultiplet. In section-\ref{scalartensor}, we will solve the hypermultiplet field equations coupled to the central charge multiplet and obtain the $\NN=2$ scalar-tensor multiplet. In section-\ref{conc}, we will end with some conclusions and future directions.
\section{$\NN=3$ conformal supergravity}\label{N3sugra}

The $\mathcal{N}=3$ conformal supergravity is a gauge theory based on the $\mathcal{N}=3$ superconformal algebra $SU(2,2|3)$, with suitable curvature constraints \cite{vanMuiden:2017qsh,Hegde:2018mxv,Hegde:2021rte}. One can obtain $\mathcal{N}=3$ Poincar{\'e} supergravity by coupling additional vector multiplet compensators to conformal supergravity and appropriately gauge fixing the extra symmetries \cite{Hegde:2022wnb}. Even before $\mathcal{N}=3$ conformal supergravity was formulated, $\mathcal{N}=3$ poincare supergravity has been studied using the group manifold apporach \cite{Castellani:1985ka}. The motivation for a superconformal apporach is that it gives a framework for constructing matter coupled $\mathcal{N}=3$ supergravity with higher derivative corrections in a systematic manner. This is possible due to the off-shell nature of the Weyl multiplet i.e. the supersymmetry algebra closes on the Weyl multiplet without the use of the equations of motion. The currently known multiplets of $\mathcal{N}=3$ conformal supergravity are the Weyl multiplet and the Vector multiplet. We briefly review these in the following subsections.

\subsection{$\mathcal{N}=3$ Weyl multiplet }

\begin{table}[t] 
		\centering
		\centering
		\begin{tabular}{|p{1cm}|p{5cm}|p{3cm}|p{2cm}| p{2cm} |}
			\hline
			Field&Properties&SU(3) irreps& Weyl weight($w$) & Chiral weight($c_A$)\\
			\hline
			\multicolumn{5}{|c|}{Independent Gauge fields}\\
			\hline
			$e_\mu^a$&Vielbein&\bf{1}&$-1$&0\\
			$\psi_\mu^I$&$\gamma_5 \psi_\mu^I=\psi_\mu^I$ &\bf{3}&$-1/2$&$-1/2$\\
			$V_\mu^{I}{}_J $&$(V_\mu^I{}_J)^*\equiv V_{\mu I}{}^J=-V_\mu{}^J{}_I $ $SU(3)_R$ gauge field &\bf{8}&0&0\\
			$A_\mu$&$ U(1)_R $gauge field&\bf{1}&0&0
            \\
			$b_\mu$&Dilatation gauge field&\bf{1}&0&0
            \\
			\hline
			\multicolumn{5}{|c|}{Dependent Gauge fields}\\
			\hline
			$\omega_{\mu}{}^{ab}$& $\omega_{\mu}{}^{ab}=-\omega_{\mu}{}^{ba}$, Spin Connection&$\bf{1}$&0&0\\
			$\phi_{\mu}^I$&$\gamma_5\phi_\mu^I=-\phi_\mu^I $&$\bf{3}$&$1/2$&$1/2$\\
			$f_\mu^a$&SCT Gauge Field&$\bf{1}$&$1$&0\\
			\hline
			\multicolumn{5}{|c|}{Covariant matter fields}\\
			\hline
 $  T^I_{ab}$& Self-dual i.e $T^I_{ab}=\frac{1}{2}\varepsilon_{abcd}T^{Icd} $&\bf{3}&$1$&$1$\\
 $E_I$& Complex & $\bf{\Bar{3}}$ &$1$&$-1$\\
$ D^I{}_J$& $(D^I{}_J)^*\equiv D_I{}^J=D^J{}_I $&\bf{8}&$2$&0\\
$\chi_{IJ}$&$\gamma_5\chi_{IJ}=\chi_{IJ} $& $\bf{\Bar{6}}$&$3/2$&$-1/2$\\
$\zeta^I$ & $\gamma_5\zeta^I=\zeta^I $&\bf{3}& $3/2$&$-1/2$\\
$\Lambda_L$ &$\gamma_5\Lambda_L=\Lambda_L$&\bf{1}&$1/2$&$-3/2$\\
 
			\hline
		\end{tabular}
		\caption{Field content of the $\mathcal{N}=3$ standard Weyl multiplet}
		\label{standard4d}	
	\end{table}
The Weyl multiplet is the multiplet that contains all the gauge fields of the superconformal algebra which are augmented by some covariant matter fields necessary to ensure the off-shell closure of the algebra. Hence the Weyl multiplet is an off-shell multiplet i.e. the superconformal algebra closes on the multiplet without the use of any field equations. The rigid superconformal algebra, however, gets modified where the structure constants are elevated to field dependent structure functions. This modified algebra is referred to as soft algebra. The $\mathcal{N}$ = 3 Weyl multiplet in four dimensions was constructed
in \cite{vanMuiden:2017qsh,Hegde:2018mxv}. It contains 64 bosonic and 64 fermionic off-shell degrees of freedom. The field contents of the Weyl multiplet along with their various properties are presented in Table \eqref{standard4d}. 

There are two kinds of supersymmetry in conformal supergravity: an ordinary supersymmetry (or $Q$-supersymmetry) and a special supersymmetry (or $S$-supersymmetry). The $Q$ and $S$-supersymmetry transformations of the fields of the Weyl multiplet parametrized by Majorana spinors $\epsilon^{I}$ and $\eta^{I}$ ($I=1,2,3$ are the $SU(3)$ R-symmetry indices) respectively are as follows:
\allowdisplaybreaks
\begin{subequations}\label{weyl}
\begin{align} 
    \delta e_{\mu}^{a}&= \bar{\epsilon}_{I}\gamma^{a}\psi_{\mu}^{I}+\text{h.c}\;,  \\
	\delta \psi_{\mu}^{I}&=2\mathcal{D}_{\mu}\epsilon^{I}-\frac{1}{8}\varepsilon^{IJK}\gamma\cdot T_{J}\gamma_{\mu}\epsilon_{K}-\varepsilon^{IJK}\bar{\epsilon}_{J}\psi_{\mu K}\Lambda_{L}-\gamma_{\mu}\eta^{I}\;,  \\
 \delta V_\mu{}^I{}_J &=\bar{\epsilon}^I\phi_{\mu J}- \frac{1}{48}\bar{\epsilon}^I\gamma_\mu\zeta_J+ \frac{1}{16}\varepsilon_{JKM}\bar{\epsilon}^K\gamma_\mu\chi^{IM}- \frac{1}{16}\bar{\epsilon}^I\gamma\cdot T_J \gamma_\mu\Lambda_R- \frac{1}{16}\bar{\epsilon}^I\gamma_\mu \Lambda_R E_J  \nonumber \\
	&+\frac{1}{8}\varepsilon_{KMJ}E^I\bar{\epsilon}^K\psi_\mu^M + \frac{1}{4}\bar{\epsilon}^I\gamma^a\psi_{\mu J}\bar{\Lambda}_L\gamma_a\Lambda_R-\bar{\psi}_\mu^I\eta_J-\text{h.c.}-\text{trace}\;,  \\
 \delta A_\mu &=\frac{i}{6}\bar{\epsilon}^I\phi_{\mu I}+ \frac{i}{36}\bar{\epsilon}^I\gamma_\mu\zeta_I+ \frac{i}{12}\varepsilon_{KMP}E^P\bar{\epsilon}^K\psi_{\mu}^M+ \frac{i}{12}\bar{\epsilon}^I\gamma\cdot T_I\gamma_\mu\Lambda_R+\frac{i}{12}\bar{\epsilon}^I\gamma_\mu\Lambda_R E_I\nonumber\\
	&-\frac{i}{3}\bar{\epsilon}^I\gamma^a\psi_{\mu I}\bar{\Lambda}_L\gamma_a\Lambda_R-\frac{i}{6}\bar{\psi}_\mu^I\eta_I+\text{h.c.
}\;, \\
\delta b_\mu &= \frac{1}{2}(\bar{\epsilon}^I\phi_{\mu I}-\bar{\psi}_\mu^I\eta_I)+\text{h.c}\;,  \\
	\delta \Lambda_L&=-\frac{1}{4}E_I\epsilon^I+\frac{1}{4}\gamma\cdot T_I\epsilon^I\;, \\
	\delta E_I &=-4 \bar{\epsilon}_I\cancel{D}\Lambda_L-\frac{1}{2}\varepsilon_{IJK}\bar{\epsilon}^J\zeta^K+\frac{1}{2}\bar{\epsilon}^J\chi_{IJ}-\frac{1}{2}\varepsilon_{IJK}E^K\bar{\epsilon}^J\Lambda_L-4\bar{\Lambda}_L\Lambda_L\bar{\epsilon}_I \Lambda_R- 4\bar{\eta}_I\Lambda_L\;, \\
	\delta T^I_{ab} &= -\bar{\epsilon}^I \cancel{D}\gamma_{ab}\Lambda_R-4\varepsilon^{IJK}\bar{\epsilon}_J R_{ab}(Q)_K+\frac{1}{8}\bar{\epsilon}_J \gamma_{ab}\chi^{IJ}+\frac{1}{24}\varepsilon^{IJK}\bar{\epsilon}_J\gamma_{ab}\zeta_K\nonumber \\
	&-\frac{1}{8}\varepsilon^{IJK}E_J\bar{\epsilon}_K\gamma_{ab}\Lambda_R +\bar{\eta}^I\gamma_{ab}\Lambda_R\;, \\
	\delta \chi_{IJ}&=2\cancel{D}E_{(I}\epsilon_{J)}-8\varepsilon_{KM(I}\gamma\cdot R(V)^M{}_{J)}\epsilon^K-2\gamma\cdot\cancel{D}T_{(I}\epsilon_{J)}+\frac{1}{3}\varepsilon_{KM(I}D^M{}_{J)}\epsilon^K\nonumber \\
 &+\frac{1}{4}\varepsilon_{KM(I}E^K\gamma\cdot T_{J)}\epsilon^M-\frac{1}{3}\bar{\Lambda
	}_L\gamma_a\epsilon_{(I}\gamma^a\zeta_{J)}+\frac{1}{4}\varepsilon_{KM(I}E_{J)}E^M\epsilon^K-\bar{\Lambda}_L\gamma^a\Lambda_R\gamma_aE_{(I}\epsilon_{J)}\nonumber\\
	&-\bar{\Lambda}_L\gamma\cdot T_{(I}\gamma^a\Lambda_R\gamma_a\epsilon_{J)}+ 2\gamma\cdot T_{(I}\eta_{J)}+ 2E_{(I}\eta_{J)}\;, \\
	\delta \zeta^I &=- 3\varepsilon^{IJK}\cancel{D}E_J\epsilon_K +\varepsilon^{IJK}\gamma\cdot\cancel{D}T_K\epsilon_J-4\gamma\cdot R(V)^I{}_J\epsilon^J-16i\gamma\cdot R(A)\epsilon^I-\frac{1}{2}D^I{}_J\epsilon^J\nonumber\\
	&-\frac{3}{8}E^I\gamma\cdot T_J\epsilon^J +\frac{3}{8}E^J\gamma\cdot T_J\epsilon^I+\frac{3}{8}E^I E_J\epsilon^J+\frac{1}{8}E^J E_J\epsilon^I\nonumber\\
	&- 4 \bar{\Lambda}_L\cancel{D}\Lambda_{R}\epsilon^I- 4 \bar{\Lambda}_R\cancel{D}\Lambda_L\epsilon^I- 3\bar{\Lambda}_R\cancel{D}\gamma_{ab}\Lambda_L\gamma^{ab}\epsilon^I-3\bar{\Lambda}_L\gamma_{ab}\cancel{D}\Lambda_R\gamma^{ab}\epsilon^I\nonumber\\
	&+\frac{1}{2}\varepsilon^{IJK}\bar{\Lambda}_L\gamma^a\epsilon_J\gamma_a\zeta_K-6\bar{\Lambda}_L\Lambda_L\bar{\Lambda}_R\Lambda_R\epsilon^I+\varepsilon^{IJK}\gamma\cdot T_J\eta_K-3\varepsilon^{IJK}E_J\eta_K\;, \\
	\delta D^I{}_J&=-3\bar{\epsilon}^I\cancel{D}\zeta_J-3\varepsilon_{JKM}\bar{\epsilon}^K\cancel{D}\chi^{IM}+\frac{1}{4}\varepsilon_{JKM}\bar{\epsilon}^I\zeta^K E^M+\frac{1}{2}\varepsilon_{JKM}\bar{\epsilon}^K\zeta^M E^I+\frac{3}{4}\bar{\epsilon}^I\chi_{JK}E^K\nonumber\\
	&+ 3\bar{\epsilon}^I\gamma\cdot T_J\overset{\leftrightarrow}{\cancel{D}}\Lambda_R-3\bar{\epsilon}^I\cancel{D}\Lambda_RE_J-3\bar{\epsilon}^I\cancel{D}E_J\Lambda_R+ \frac{3}{4}\varepsilon_{JKM}E^M\bar{\epsilon}^K\Lambda_L E^I\nonumber\\
	&+ {3\varepsilon_{JKM}T^I\cdot T^M\bar{\epsilon}^K\Lambda_{L}}-2\bar{\epsilon}^I\Lambda_L\bar{\Lambda}_R\zeta_J-3\bar{\epsilon}^I\Lambda_L\bar{\Lambda}_R\Lambda_RE_J+3\bar{\epsilon}^I\gamma\cdot T_J\Lambda_L\bar{\Lambda}_R\Lambda_R\nonumber \\
    &+\text{h.c.} -\text{trace}\;,
	\end{align}
\end{subequations}

where, $\mathcal{D}_{\mu}\epsilon^{I}$ is defined as:
	\begin{align}\label{Depsilon}
\mathcal{D}_\mu\epsilon^I=\partial_\mu\epsilon^I-\frac{1}{4}\gamma\cdot\omega_\mu\epsilon^I+\frac{1}{2}(b_\mu+iA_\mu)\epsilon^I-V_\mu{}^I{}_J\epsilon^J\;. 
	\end{align}

\textbf{Notations:} In four dimensional conformal supergravity, chiral notation is followed throughout the literature, which we also follow here. In this notation, raising and lowering of the R-symmetry indices $I,J,\cdots$ is via complex conjugation. For example, the field $T_{ab}^{I}$, which is the self dual component of a real anti-symmetric tensor field will go to $T_{abI}$ under complex conjugation which is the anti-self dual component of the same tensor field i.e. $T_{abI}=(T_{ab}^{I})^*$. For bosonic fields such as $E_{I}$ which are not real, complex conjugation will also change the position of the R-symmetry indices in similar fashion i.e $E^{I}=\left(E_I\right)^*$. Similarly for fermions, which are represented by Majorana spinors, complex conjugation will act in the same way by changing the position of the R-symmetry indices. For instance, the spinor $\zeta^{I}$ which is the left chiral part of a Majorana spinor, will go to $\zeta_{I}$ under complex conjugation, which is the right chiral part of the same Majorana spinor. Precisely:
\begin{align}
    \zeta_{I}=i\gamma^0c^{-1}\left(\zeta^I\right)^*\;,
\end{align}
where $c$ is the charge conjugation matrix in four dimensions. The subscripts $L$, $R$ on the fields $\Lambda_L, \Lambda_R$ denote the left and right chirality respectively and are not to be confused as an $SU(3)$ index. $L$ and $R$ are never used as $SU(3)$ indices in this paper.

\subsection{$\mathcal{N}$=3 Vector Multiplet}
The $\mathcal{N}=3$ vector multiplet coupled to the standard Weyl multiplet was constructed in \cite{Hegde:2022wnb} through a supersymmetric truncation of the $\mathcal{N}$ = 4 vector multiplet. This multiplet is an on-shell multiplet i.e. the supersymmetry algebra closes on this multiplet upon using the field equations. It contains 8+8 on-shell degrees of freedom. The field content of this multiplet is given in Table \eqref{vect table} along with their properties. 
\begin{table}[h!]
	\centering
	\centering
	\begin{tabular}{ |p{1.5cm}|p{1.5cm}|p{2cm}|p{1cm}|p{2cm}|p{2cm}| }
		\hline
		Field & Type& $SU(3)$ irreps &$w$ (Weyl weight) &$c_A$ (chiral weight)\\
		\hline
		$C_\mu$& Boson&\bf{1}&0&0\\
		$\xi_I$&Boson&$\bf{\bar{3}}$&$1$&$-1$\\
		$\psi_{I}$&{Fermion}&$\bf{\bar{3}}$&$3/2$&$1/2$\\
  $\theta_L$ & Fermion & \bf{1}& $3/2$& $3/2$\\
		\hline
	\end{tabular}
	\caption{$\NN=3$ Vector Multiplet}\label{vect table}	
\end{table}

The $Q$ as well as $S$ Supersymmetry transformations of the components of the $\mathcal{\mathcal{N}}$=3 vector multiplet are as follows;\\
 \begin{align} \label{n=3 vector}
\delta{C}_{\mu}&= \bar{\epsilon}^{I}\gamma_{\mu}\psi_{I}-2\bar{\epsilon}^{I}\psi_{\mu}^J\xi^K \varepsilon_{IJK} - \bar{\epsilon}_{I}\gamma_{\mu}\Lambda_L \xi^I +\text{h.c}\;, \nonumber \\
\delta \psi_{I}&= -\frac{1}{2}   \gamma \cdot \mathcal{F}^+ \epsilon_I - 2 \varepsilon_{IJK} \cancel{D} \xi^K \epsilon^J - \frac{1}{4} E_I \xi^K \epsilon_K + \frac{1}{2} \bar{\Lambda}_L \theta_L \epsilon_I\nonumber \\
& + \frac{1}{2}\gamma_a \epsilon^J \bar{\Lambda}_R \gamma^a \Lambda_L \xi^K \varepsilon_{IJK} + 2 \varepsilon_{IJK} \xi^J \eta^K \;,
\nonumber \\
\delta \theta_L &= - 2 \cancel{D} \xi^I \epsilon_I -  \gamma^a \bar{\Lambda}_L \gamma_a \Lambda_R \xi^J \epsilon_J + \frac{1}{4} \varepsilon_{IJK} E^I \xi^J \epsilon^K - \bar{\Lambda}_R \psi_I \epsilon^I  - 2 \xi^I \eta_I\;, 
\nonumber \\
\delta \xi_I&=-\bar{\epsilon}_I \theta_R + \varepsilon_{IJK} \bar{\epsilon}^J \psi^K\;,  \nonumber \\
\end{align}
where, $\mathcal{F}_{a b}^{+}$ is the self-dual part of the modified superconformal field strength of $C_\mu$ defined as follows. 

\begin{align} \label{mod F}
      \mathcal{F}_{a b}^{+}= &\widehat{F}_{a b}^{+}-\frac{1}{4} \bar{\Lambda}_R \gamma_{a b} \theta_R-\frac{1}{2} T_{a b}^I \xi_I\;,
\end{align}
where $\hat{F}_{ab}$ is the un-modified supercovariant field strength associated with the vector field $C_\mu$ given as:
\begin{align}\label{F}
   \widehat{F}_{ab} =&  2 e_{[a}^\mu e_{b]}^\nu \partial_\mu C_\nu - \left( \bar{\psi}^I_{[a} \gamma_{b]} \psi_I - \epsilon_{IJK} \bar{\psi}_a^I \psi_b^J \xi^K - \bar{\psi}_I {}_{[a} \gamma_{b]} \Lambda_L \xi^I + \text{h.c}\right)\;.
   \end{align}
Being an on-shell multiplet, the superconformal algebra closes on the vector multiplet fields only when their field equations are satisfied which are given below,
\begin{subequations}\label{n=3 vector eom}
\begin{align} 
& \cancel{ D} \psi_I+\frac{1}{2} \bar{\Lambda}_R \psi_I \Lambda_L-\frac{1}{8} E_I \theta_L+\frac{1}{8} \gamma \cdot T_I \theta_L+\frac{1}{8} \chi_{IJ} \xi^J+\frac{1}{24} \varepsilon_{IJK} \zeta^J \xi^K=0\;, \label{psieom} \\[5pt] 
& \cancel{ D} \theta_R-\frac{3}{4} \bar{\Lambda}_R \theta_R \Lambda_L+\frac{1}{4} \gamma \cdot \hat{F}^{-} \Lambda_L-\frac{3}{8} \bar{\Lambda}_L \Lambda_L \theta_L-\frac{1}{8} \gamma \cdot T_I \Lambda_L \xi^I-\frac{1}{8} E_I \psi^I-\frac{1}{8} \gamma \cdot T_I \psi^I \nonumber \\ 
& -\frac{1}{12} \zeta^I \xi_I-\frac{1}{8} E^I \xi_I \Lambda_L=0\;, \label{thetaeom} \\[5pt]
&\square_c \xi_J+ \frac{1}{4} D_a (\bar{\Lambda}_R \gamma_a \Lambda_L \xi_J ) +\frac{1}{4} \bar{\Lambda}_R \cancel{ D} \xi_J \Lambda_L+\frac{1}{4} \widehat{F}^{-} \cdot T_J-\frac{1}{16} \bar{\Lambda}_L \gamma \cdot T_J \theta_L-\frac{1}{8} \xi^I T_I \cdot T_J+\frac{1}{24} \bar{\zeta}_J \theta_R \nonumber\\
& +\frac{1}{16} E_J \bar{\Lambda}_R \theta_R -\frac{1}{16} \bar{\chi}_{M J} \psi^M-\frac{1}{48} \varepsilon_{JKM} \bar{\zeta}^M \psi^K-\frac{1}{48} D^K{}_J \xi_K-\frac{1}{96} \xi_J E^K E_K+\frac{1}{12} \xi_J \left(\bar{\Lambda}_R \cancel{ D} \Lambda_L+\bar{\Lambda}_L \cancel{ D} \Lambda_R\right) \nonumber \\
& +\frac{1}{12} \xi_J \bar{\Lambda}_R \Lambda_R \bar{\Lambda}_L \Lambda_L=0\;, \label{xieom}  \\[5pt]
&D_a\left({G}^{+a b}-{G}^{-a b}\right)=0\;, \label{maxeom}
\end{align}
\end{subequations}
where $G^{+}_{ab}$ is defined as:
\begin{align}
   {G}_{ab}^+=& - i \widehat{F}_{ab}^+ +\frac{i}{2} \bar{\Lambda}_R \gamma_{ab} \theta_R + iT_{ab}^I \xi_I\;,
\end{align}
and $G_{ab}^{-}$ is the hermitian conjugate of $G^{+}_{ab}$.

\section{Supersymmetric Truncation}\label{susytrunc} 
Supersymmetric truncation is a procedure that reduces an $\mathcal{N}$ - extended conformal supergravity to $(\mathcal{N}-1)$-extended conformal supergravity. In supersymmetric truncation of an $\mathcal{N}$- extended conformal supergravity, we set the $\mathcal{N}^{th}$ supersymmetry parameter ($\epsilon^\mathcal{N}$ ) and the corresponding gravitino ($\psi_\mu^\mathcal{N}$) to zero. The condition $\psi_\mu^\mathcal{N}$ =0 is not supersymmetric; thus, sequentially applying supersymmetry transformations will give us further conditions that several other field components must also vanish. In addition, the extra gauge symmetries of $\mathcal{N}$ - extended theory compared to $\mathcal{N} - 1$ extended theory will also be broken. Thus, the resulting theory will be an $\mathcal{N}-1$ extended conformal supergravity and the non-vanishing fields can be rearranged into $\mathcal{N}-1$ multiplets. The matter multiplets of $\mathcal{N}$-extended theory also reduce to those of the $\mathcal{N}-1$ theory in the same way. It is already known that supersymmetric truncation of the $\mathcal{N}=4$ Weyl multiplet and vector multiplet gives the $\mathcal{N}=3$ Weyl multiplet and vector multiplet respectively \cite{Hegde:2022wnb}. Similarly, the truncation of the $\mathcal{N}=2$ Weyl multiplet gives the $\mathcal{N}=1$ Weyl multiplet and the $\mathcal{N}=1$ Vector multiplet \cite{Yamada:2019ttz}. In this section we discuss the supersymmetric truncation of $\mathcal{N}=3$ Weyl multiplet and vector multiplet.
\subsection{Supersymmetric Truncation of $\mathcal{N}=3$ Weyl multiplet} \label{truncation}
As mentioned above, we start by setting the third supersymmetry parameter and the third gravitino to zero.  
\begin{equation}
\epsilon^3 = 0 =\psi^3_\mu\;.
\end{equation}
Using \eqref{weyl} this will lead to a set of truncations, as shown below in \eqref{truncated fields}. The $SU(3) \times U(1)$ R-Symmetry group of $\mathcal{N} = 3$ Weyl multiplet breaks into $SU(2) \times U(1) \times U(1)$ out of which an $SU(2)\times U(1)$ becomes the R-symmetry of the $\mathcal{N}=2$ theory and the other $U(1)$ becomes a gauge symmetry associated with a vector multiplet. The $SU(3)$ indices $I,J,K\cdots\in \{1,2,3\}$ decompose to $I=3$ and $SU(2)$ indices $i,j,k\cdots\in$ $\{1,2\}$.

\begin{align} \label{truncated fields}
    \psi_{\mu}^3=0 \implies  & \Lambda_L =0\;, \; T_{ab}^{i} =0\; \text{  and  }\; V_{\mu}{}^{i}{}_3 = 0\;. \nonumber \\
    V_{\mu}{}^{i}{}_{3}=0 \implies & \phi_\mu^3 = 0\;, \; \zeta^3 = 0\;, \; \chi^{ij} = 0\;, \; \chi^{33} = 0\; \text{  and  }\; E^i=0\;. \nonumber\\
    \chi_{ij} = 0 \implies & D^{3}{}_j = 0\;.
\end{align}
The S-supersymmetry and $SU(3)$ parameters are also restricted as ;
\begin{align}
    \eta^3=0\;,\;
    \Lambda^{i}{}_3=0\;. \nonumber
\end{align} 
The non vanishing fields $i.e,$ 
($e_\mu^a$, $\psi_\mu ^i$, $V_{\mu}{}^{i}{}_{j}$, $V_{\mu}{}^{3}{}_{3}$, $A_\mu$, $b_\mu$, $E_3$, $T_{ab3}$, $\chi_{i3}$, $\zeta_i$, $D^{i}{}_j, D^{3}{}_3)$
rearrange themselves into the $\mathcal{N}=2$ Weyl multiplet and $\mathcal{N}=2$ vector multiplet, as per the dictionaries given in the next subsections. 
\subsubsection{$\mathcal{N}=2$ Weyl multiplet} \label{N=2 weyl}
The following combinations of fields form the off-shell $\mathcal{N}=2$ Weyl multiplet. Here fields on the LHS are $\mathcal{N}=2$ fields and those in the RHS are $\mathcal{N}=3$ fields.

\begin{equation} \label{dict weyl-weyl}
    \begin{split}
        & e^a_{\mu}= e^a_{\mu} \; , \\
        &\psi^i_{\mu}=  \psi^i_{\mu} \; ,\\
        &b_\mu = b_\mu \; ,\\
        &V_\mu{}^i{}_j = V_\mu{}^i{}_j + \frac{1}{2} \delta^i_j V_\mu{}^3{}_3 \; , \\  
         &T^{-}_{ab}= - T_{ab3}\; ,\\
        &A_\mu = A_\mu - iV_\mu{}^3{}_3 \; ,\\
        &D=\frac{1}{48}D^3{}_3 + \frac{1}{96} E^3E_3 \; , \\
        &\chi_i = - \frac{1}{16} \epsilon_{ij} \chi^{j3} + \frac{1}{48} \zeta_i \; .\\ 
        \end{split}       
\end{equation}

Their supersymmetry transformations can be obtained by substituting the truncation conditions \eqref{truncated fields} into  the transformations \eqref{weyl}. The transformations of the $\mathcal{N}=2$ Weyl multiplet is given below, 
\begin{equation}
\begin{aligned}
\delta e_\mu{ }= & \bar{\epsilon}^i \gamma^a \psi_{\mu i}+\text { h.c } \; , \\
\delta \psi_\mu^i= & 2 \mathcal{D}_\mu \epsilon^i-\frac{1}{8} \varepsilon^{ij} \gamma \cdot T^{ -} \gamma_\mu \epsilon_j - \gamma_\mu \eta^i \; , \\
\delta b_\mu= & \frac{1}{2} \bar{\epsilon}^i \phi_{\mu i} - \frac{1}{2} \bar{\psi}_{\mu}^i \eta_i+\text { h.c } \; ,\\
\delta A_\mu= & \frac{i}{2}  \bar{\epsilon}^i \phi_{\mu i}+ i \bar{\epsilon}^i \gamma_\mu \chi_i - \frac{i}{2} \bar{\psi}_{\mu}^i \eta_i +\text { h.c } \; , \\
\delta V_{\mu j}^i= &  \bar{\epsilon}^i \phi_{\mu j}- \bar{\epsilon}^i \gamma_\mu \chi_j + \bar{\eta}^i \psi_{\mu j} -\text{h.c - trace} \; ,  \\
\delta T_{a b}^{-}= & 4 \varepsilon_{ij}  \bar{\epsilon}^{i} R_{a b}(Q)^{j} -2 \varepsilon_{ij} \epsilon^{i} \gamma_{ab} \chi^{j} \; , \\
\delta \chi^i= & -\frac{1}{12} \varepsilon^{ij} \gamma_{ab} \cancel D T^{a b -} \epsilon_j -\frac{1}{3} \gamma \cdot R(V)^i{ }_{j}  \epsilon^j-\frac{i}{3}  \gamma \cdot R(A)  \epsilon^i \;  \\
& +D \epsilon^i+\frac{1}{12} \varepsilon^{ij} \gamma \cdot T^{-} \eta_j \; ,\\
\delta D= & \bar{\epsilon}^i \cancel D \chi_i+\text { h.c} \; .
\end{aligned}
\end{equation}

The $\mathcal{N} =2$ Weyl multiplet obtained in this section matches with the literature except for some field redefinitions and a field dependent special conformal transformation, as shown in appendix \ref{conventions}. 

\subsubsection{$\mathcal{N}=2$ Vector Multiplet}
The remaining non-vanishing fields form the off-shell $\mathcal{N}=2$ Vector multiplet as per the following dictionary ($\mathcal{N}=2$ fields on LHS and $\mathcal{N}=3$ fields on RHS);

    \begin{align} \label{dict weyl-vector}
         W_\mu &= i V_\mu ^3{}_3 +2 A_\mu \; , \nonumber\\
        X &=- \frac{i}{8}E_3 \; , \nonumber  \\
        \Omega_i &= -\frac{i}{16} \chi_{i3}+ \frac{i}{16} \epsilon_{ij} \zeta^j \; , \nonumber \\
        Y_{ij} &=\frac{i}{24} \varepsilon_{k(i}D^k{}_{j)} \; .
    \end{align}
The auxiliarly field $Y_{ij}$ satisfies the psuedo reality condition as required by the $\NN=2$ vector multiplet, 
\begin{align} \label{psuedoY}
    Y_{ij}= \epsilon_{ik} \epsilon_{jl} Y^{kl} \; .
\end{align}

The supersymmetry transformations of the off-shell $\mathcal{N}=2$ vector multiplet are given below,

\begin{align} \label{N=2 vector}
\delta X & =\bar{\epsilon}^i \Omega_i \; , \nonumber \\
\delta \Omega_i & =2 \cancel D X \epsilon_i+\frac{1}{2} \varepsilon_{i j} \gamma \cdot \mathcal{F}^{-}  \epsilon^j+Y_{i j} \epsilon^j+2 X \eta_i \; , \nonumber \\
\delta W_\mu & =\varepsilon^{i j} \bar{\epsilon}_i \gamma_\mu \Omega_j+2 \varepsilon_{i j} \bar{\epsilon}^i \bar{X} \psi_\mu^j+\text { h.c } \; , \nonumber \\
\delta Y_{i j} & =2 \bar{\epsilon}_{(i} \cancel D \Omega_{j)}-4X \bar{\epsilon}_{(i}\chi_{j)} +2 \varepsilon_{i k} \varepsilon_{j l} \bar{\epsilon}^{(k} \cancel D \Omega^{l) } - 4 \varepsilon_{ik}  \varepsilon_{jl} \bar{X} \bar{\epsilon}^{(k} \chi^{l)} \; ,
\end{align}
where $\mathcal{F}_{\mu \nu}$ is the modified superconformally covariant field strength of the gauge field $W_\mu$.
\begin{align}
    \mathcal{F}_{\mu \nu}=2 \partial_{[\mu} W_{\nu]}-\varepsilon_{i j} \bar{\psi}_{[\mu}{ }^i\left(\gamma_{\nu]} \Omega^j+\psi_{\nu]}{ }^j \bar{X}\right)-\varepsilon^{i j} \bar{\psi}_{[\mu i}\left(\gamma_{\nu]} \Omega_j+\psi_{\nu] j} X\right) -\frac{1}{2} \bar{X} T_{\mu \nu}^- -\frac{1}{2} X T_{\mu \nu}^+ \; .
\end{align}

The above transformations matches with those given in \cite{Butter:2017pbp}. Since, this $\NN=2$ vector multiplet arises from the supersymmetric truncation of the $\NN=3$ Weyl multiplet, we can interpret it as a compensating vector multiplet that is necessary to go from $\NN=2$ conformal supergravity to Poincar{\'e} supergravity. The compensating vector multiplet provides a gauge field that can be interpreted as a graviphoton in the Poincar{\'e} theory, which is the gauge field corresponding to the central charge generator in the super-Poincar{\'e} algebra. Hence we will refer to this $\NN=2$ vector multiplet as the central charge multiplet and the corresponding $U(1)$ as the central charge transformation.

\subsection{Truncation of $\mathcal{N}$=3 Vector multiplet}
We perform supersymmetric truncation of the on-shell $\mathcal{N}=3$ Vector multiplet in a similar way as we truncated the Weyl multiplet in the previous subsection. It is easily seen that under this truncation, the $\mathcal{N}=3$ vector multiplet is reduced to two $\mathcal{N}=2$ irreducible multiplets, the $\mathcal{N}=2$ on-shell vector multiplet and an on-shell scalar multiplet. \\

The following fields go into the on-shell vector multiplet. We have $\mathcal{N}=2$ fields on the LHS and $\mathcal{N}=3$ fields on the RHS. 
\begin{align} \label{vector to vector}
    C_\mu & := C_\mu \; , \nonumber\\
    \psi_i & := \psi_i \; , \nonumber\\
     \xi& : = \xi_3 \; . \nonumber\\
\end{align}

As can be seen from the above identification of the $\NN=2$ vector multiplet, we do not get anything corresponding to the auxiliary field $Y_{ij}$ that appears in an off-shell vector multiplet. Hence, the $\NN=2$ vector mutiplet that we obtain from the supersymmetric truncation of an on-shell $\NN=3$ vector multiplet is an on-shell multiplet as opposed to the $\NN=2$ vector multiplet that we obtain from the supersymmetric truncation of the $\NN=3$ Weyl multiplet which is off-shell. 

The remaining fields form an on-shell $\NN=2$ scalar multiplet. In the L.H.S, we have $\mathcal{N}=2$ fields and on the RHS we have $\mathcal{N}=3$ fields. 
\begin{align} \label{vector-hyper}
    \psi_R& := \psi_3 \; , \nonumber \\ 
    \theta_R& := \theta_R \; , \nonumber \\ 
    \xi_i &:= \xi_i \; .
\end{align}
This scalar multiplet, as we will show in the next section constitutes an on-shell $\NN=2$ hypermultiplet with a broken rigid $SU(2)$ and a non-trivial central charge transformation, thereby constituting a massive hypermultiplet. We set the on-shell vector multiplet to zero and focus on this hypermultiplet.
Their supersymmetry transformations and equations of motion can be obtained from supersymmetric truncation of the parent $\mathcal{N}=3$ vector multiplet. Their supersymmetry transformations are as follows, 

 \begin{align} \label{hyper transform}
\delta \psi_{R}&= - 2 \varepsilon_{jk} \cancel{D} \xi^k \epsilon^j - 2i X \xi^k \epsilon_k + 2 \varepsilon_{jk} \xi^j \eta^k\;, 
\nonumber \\
\delta \theta_L &= - 2 \cancel{D} \xi^i \epsilon_i  - 2i \varepsilon_{jk} \bar{X} \xi^j \epsilon^k  - 2 \xi^i \eta_i\;, 
\nonumber \\
\delta \xi_i&=-\bar{\epsilon_i} \theta_R + \varepsilon_{ij} \bar{\epsilon}^j \psi_L\;. \nonumber \\
\end{align}
Their field equations are as follows,

\begin{subequations}\label{eom n=2 hyper}
\begin{align} 
& \cancel{D} \psi_R-i X \theta_L-\frac{1}{8} \gamma \cdot T^- \theta_L + \varepsilon_{ij} \chi^j \xi^i + i \Omega_j \xi^j = 0\;,  \label{psieq}  \\
& \cancel{D} \theta_R-i X \psi_L +\frac{1}{8} \gamma \cdot T^- \psi_L - \chi^i \xi_i - i \varepsilon^{ij} \Omega_j \xi_i=0\;, \label{thetaeq} \\
&\square_c \xi_i - \frac{i}{2} \varepsilon_{ij} \bar{\Omega}^j \theta_R - \frac{i}{2} \bar{\Omega}_i \psi_L +\frac{1}{2} \bar{\chi}_i \theta_R  - \frac{1}{2} \varepsilon_{ij} \bar{\chi}^j \psi_L + \frac{1}{2}D \xi_i - \frac{i}{2}\varepsilon_{ik} Y^{kj} \xi_j -  \lvert X \rvert ^2 \xi_i =0\;, \label{xieq} 
\end{align}
\end{subequations}

where the dictionaries \eqref{dict weyl-weyl}, \eqref{dict weyl-vector} and \eqref{vector-hyper} are used. Since this hypermultiplet is charged under the central charge transformation, one will find the fields of the central charge multiplet appearing in its supersymmetry transformations as well as field equations. We will make use of this to construct a new $\mathcal{N}=2$ multiplet in section \ref{scalartensor}. We would also like to point out that the above multiplet can also be obtained from a double supersymmetric truncation of the $\NN=4$ vector multiplet as we have explicitly seen. We refrain from giving the details for the sake of brevity.

\section{Off-shell $\NN=2$ hypermultiplet and its relation to the $\NN=2$ hypermultiplet obtained above}\label{N2hyper}

In this section we will give a brief review of the $\mathcal{N} =2$ off-shell hypermultiplet with a central charge. One may refer to \cite{deWit:1980lyi,Kleijn:1998zea} for detailed descriptions. The physical degrees of freedom are scalar field $A_{\alpha}{}^{i}$ which constitutes a doublet under the $SU(2)$ R-symmetry and Majorana spinors $\zeta^\alpha$ which are singlet under the $SU(2)$ R-symmetry. These fields also transform under a rigid $SU(2)$ with indices $\alpha, \beta \in \{\underline{1}, \underline{2} \}  $. The scalar field satisfies a pseudo reality condition, 

\begin{align}\label{real_cond}
    A_{\alpha}{}^i = (A^{\alpha}{}_i)^*= \varepsilon^{ij} \varepsilon_{\alpha \beta} A^{\beta}{}_{j} \; . 
\end{align}

The chirality of spinors is encoded in the position of the rigid $SU(2)$ index.

\begin{align}
    \zeta^{\alpha}= -\gamma_5 \zeta^{\alpha} \quad, \quad \zeta_{\alpha}= \gamma_5 \zeta_{\alpha} \; .
\end{align}

The off-shell hypermultiplet has a central charge transformation (denoted by $\delta_z$), successive application of which generates an infinite series of fields denoted by ( $A^{(z)}, \zeta^{(z)} $), ( $A^{(zz)}, \zeta^{(zz)} $).. etc as shown below,

\begin{align}\label{cent_charge_trs}
   \delta_z(z) A_{\alpha}{}^i =\frac{z}{2}A^{(z)}_{\alpha}{}^{i}\;, \; \delta_z (z) \zeta^{\alpha} =\frac{z}{2} \zeta^{(z) \alpha } \;, \;\delta_z(z) A^{(z)}_{\alpha}{}^{i} = \frac{z}{2}A^{(zz)}_{\alpha}{}^{i}\;, \;\delta_z (z) \zeta^{(z)\alpha} =\frac{z}{2} \zeta^{(zz) \alpha }\; \cdots
\end{align}

The supersymmetry transformations of the hypermultiplet fields are given below,

\begin{align} \label{off hyp transform}
    \delta A^{\alpha}{}_{i} &= 2  \bar{\epsilon}_i \zeta^{\alpha}+ 2 \epsilon^{\alpha \beta} \varepsilon_{ij} \bar{\epsilon}^j \zeta_{\beta} \; , \nonumber \\
    \delta \zeta^{\alpha}&= \cancel{D}A^{\alpha}{}_{i} \epsilon^i + \varepsilon^{ij} X A^{(z) \alpha}{}_{i} \epsilon_j + A^{\alpha}{}_i \eta^i \; .  
\end{align} 

Note that the derivatives in the above are also covariant with respect to central charge transformations. Replacing $A^{\alpha}{}_{i}$ with $A^{(z) \alpha}{}_{i}$, $\zeta^{\alpha} $ with  $\zeta^{(z) \alpha}$ and $A^{(z) \alpha}{}_{i}$ with $A^{(zz) \alpha}{}_{i}$ in \eqref{off hyp transform}, gives us the 
transformations of $A^{(z)\alpha}{}_i$ and $\zeta^{(z) \alpha}$. Similarly proceeding one can write the transformations of higher ``$z$'' fields. The superconformal algebra closes on these fields only when the following constraints are satisfied. 

\begin{align} \label{off hyp constraint}
    \epsilon^{\alpha \beta} \cancel{D} \zeta_{\beta} &+ \frac{1}{2} A^{(z) \alpha }{}_i \Omega^i + \bar{X} \zeta^{(z) \alpha} + \frac{1}{8} \gamma \cdot T^+ \zeta^{\alpha} + \frac{1}{2} \varepsilon^{ij} \chi_j A^{\alpha}{}_{i}=0  \; , \nonumber \\
     \square_c A^{\alpha}{}_i &-  \varepsilon_{ij} \bar{\Omega}^j \zeta^{(z) \alpha}+\epsilon^{\alpha \beta} \bar{\Omega}_i \zeta_\beta^{(z)} - \bar{\chi}_i \zeta^{\alpha} - \varepsilon_{ij} \epsilon^{\alpha \beta} \bar{\chi}^j \zeta_{\beta} +\frac{1}{2} \varepsilon_{ik} A^{(z) \alpha}{}_j Y^{kj} \nonumber \\
     &  + \frac{1}{2} D A^{\alpha}{}_i - \lvert X \rvert ^2 A^{(zz) \alpha}{}_i  =0  \;.  
\end{align}

Applying central charge transformations (\ref{cent_charge_trs}) on the above constraints generate an infinite sequence of other constraints. It is easy to see that from these constraints the fields $A^{(zz)\alpha{}}{}{}_i$ and $\zeta^{(z)\alpha}$ can be completely determined. Similarly, using the other constraints all the higher ``$z$'' fields can also be determined, leaving only $A^{\alpha}{}_i$, $\zeta^{\alpha}$ and $A^{(z)\alpha}{}_i$ as independent fields. \\

Let us make a specific choice of central charge transformation.

\begin{align} \label{central charge}
&\delta_z(z)A^{\underline{1}}{}_i = - \frac{iz}{2}A^{\underline{1}}{}_i  , \quad  \delta_z(z) \zeta^{\underline{1}} = -\frac{iz}{2}\zeta^{\underline{1}} \;, \nonumber \\
    &\delta_z(z)A^{\underline{2}}{}_i = \frac{iz}{2} A^{\underline{2}}{}_i  , \quad  \delta_z(z) \zeta^{\underline{2}} = \frac{iz}{2}\zeta^{\underline{2}} \;.
\end{align}

This choice does a couple of things. Firstly, it converts the off-shell hypermultiplet into an on-shell massive hypermultiplet\footnote{We call it a massive hypermultiplet because of the nontrivial central charge transformation. If one takes a trivial central charge transformation one can obtain an on-shell massless hypermultiplet \cite{Gold:2022bdk}.} where the constraints (\ref{off hyp constraint}) becomes the field equations. Secondly, it breaks the rigid $SU(2)$, since the central charge acts on each $\alpha$ component differently. Now, lets rename the fields as follows, 

\begin{align}
    \xi_i \equiv A^{\underline{1}}{}_i, \quad \theta_R \equiv -2 \zeta^{\underline{1}}, \quad \psi_L \equiv 2 \zeta_{\underline{2}} \;.
\end{align}

$A^{\underline{2}}{}_i$ is related to $A^{\underline{1}}{}_i$ by the reality condition \eqref{real_cond}. So essentially, we have an $SU(2)$ doublet of complex scalar $\xi_i$ in the bosonic sector. One can check that under this choice, the transformations \eqref{off hyp transform} matches with \eqref{hyper transform} and the constraints \eqref{off hyp constraint} matches with the equations of motion in  \eqref{eom n=2 hyper}. Thus, the scalar multiplet we discussed in the previous section is nothing but an on-shell massive hypermultiplet that breaks the rigid $SU(2)$ which can also be obtained from an off-shell hypermultiplet by making the special choice of central charge transformations given in \eqref{central charge}. 

\section{The $\mathcal{N}=2$ Scalar-Tensor multiplet}\label{scalartensor}
In this section, we interpret the field equations of the hypermultiplet (\ref{eom n=2 hyper}) as constraints on the central charge multiplet and replace some of the components of the central charge multiplet by those of the hypermultiplet and a two-form gauge field. This is similar in spirit to the construction of dilaton Weyl multiplets, where one interprets the equations of motion of an on-shell matter multiplet as constraints on the standard Weyl multiplet and uses them to replace some of the standard Weyl multiplet fields by those of the matter multiplets (See, for example \cite{Butter:2017pbp} and \cite{Gold:2022bdk} for $\NN=2$ dilaton Weyl multiplets in four dimensions, \cite{Adhikari:2024qxg} for $\NN=3$ dilaton Weyl multiplets in four dimensions, \cite{Ciceri:2024xxf} and \cite{Adhikari:2024esl} for $\NN=4$ dilaton Weyl multiplets in four dimensions, \cite{Bergshoeff:2001hc} for $\NN=1$ dilaton Weyl multiplet in five dimensions, \cite{Bergshoeff:1985mz} and \cite{Adhikari:2024esl} for dilaton Weyl multiplets in $(1,0)$ and $(2,0)$ conformal supergravity in six dimensions respectively.) 

\subsection{Interpreting equations of motion}
As will be shown in this subsection, the fields $\Omega_i, Y_{ij}$ and $W_{\mu}$ of the central charge multiplet will become composite using the field equations \eqref{eom n=2 hyper}. For this reason, we will deonte these fields with a mathring from here on, $i.e,$ as $\mathring{\Omega}_i, \mathring{Y}_{ij}$ and $\mathring{W}_\mu$.
First, we will replace the $SU(2)$ doublet spinor $\mathring{\Omega}_i$ of the central charge multiplet by the two $SU(2)$ singlet spinors $\theta$ and $\psi$ of the hypermultiplet. We have to do this in two steps. In the first step,
we use the field equation of $\psi_R$ in \eqref{psieq} to write ,

\begin{align}\label{omega_psi}
   \mathring{\Omega}_j \xi^j = i\bigg(  \cancel{D} \psi_R-i X \theta_L-\frac{1}{8} \gamma \cdot T^- \theta_L + \varepsilon_{ij} \chi^j \xi^i \bigg) \;.
\end{align}
 In the second step, we use the field equation of $\theta_R$ \eqref{thetaeq} to write,

 \begin{align}
      \varepsilon^{kl} \mathring{\Omega}_l \xi_k = -i \bigg( \cancel{D} \theta_R-i X \psi_L +\frac{1}{8} \gamma \cdot T^- \psi_L - \chi^i \xi_i \bigg) \;.
 \end{align}
 Multiplying this with $\varepsilon_{ij}$ we get,

 \begin{align}\label{omega_theta}
     \xi_i \mathring{\Omega}_j - \xi_j \mathring{\Omega}_i = -i \varepsilon_{ij} \bigg( \cancel{D} \theta_R-i X \psi_L +\frac{1}{8} \gamma \cdot T^- \psi_L - \chi^i \xi_i \bigg) \;.
 \end{align}

 Contracting \eqref{omega_theta} with $\frac{\xi^i}{ \xi^m \xi_m}$ and using \eqref{omega_psi}, we get,

 \begin{align}\label{omegasol}
     \mathring{\Omega}_i =&  \frac{i \xi_i }{\xi^m \xi_m} \bigg(  \cancel{D} \psi_R-i X \theta_L-\frac{1}{8} \gamma \cdot T^- \theta_L + \varepsilon_{kj} \chi^j \xi^k \bigg)  \nonumber \\
     & +i \varepsilon_{ij} \frac{\xi^j}{\xi^m \xi_m} \bigg( \cancel{D} \theta_R-i X \psi_L +\frac{1}{8} \gamma \cdot T^- \psi_L - \chi^k \xi_k \bigg) \;.
 \end{align}\\


 Now, we turn to the field equations of $\xi_i$ \eqref{xieq}. Since $\xi_i$ is a complex $SU(2)$ doublet of scalars, its equations have 4 components. As will be shown below, three of the components can be used to determine $\mathring{Y}_{ij}$ completely. The remaining one component can be used to trade the $U(1)$ gauge field $\mathring{W}_\mu$  of the central charge multiplet for a two-form gauge field.    The field equations \eqref{xieq} can be written as,
 \begin{align} \label{Yij}
   &\frac{i}{2}\varepsilon_{ik} \mathring{Y}^{kj} \xi_j= \Gamma_i\;,
\end{align}
where, we have defined
\begin{align}\label{Gammadef}
    \Gamma_i \equiv  \square_c \xi_i - \frac{i}{2} \varepsilon_{ij} \mathring{\bar{\Omega}}^j \theta_R - \frac{i}{2} \mathring{\bar{\Omega}}_i \psi_L +\frac{1}{2} \bar{\chi}_i \theta_R  - \frac{1}{2} \varepsilon_{ij} \bar{\chi}^j \psi_L + \frac{1}{2}D \xi_i  -  \lvert X \rvert ^2 \xi_i\;.
\end{align}
Using the psuedo reality property \eqref{psuedoY} of $\mathring{Y}_{ij}$, we can write \eqref{Yij} as  
\begin{align}\label{YG}
   -&\frac{i}{2}  \varepsilon^{jm} \mathring{Y}_{im} \xi_j = \Gamma_i  \;.
\end{align}
Multiplying the above equation \eqref{YG} with $\varepsilon_{kl} \xi^l$ we obtain, 
\begin{align}\label{YGGamma}
     &\frac{i}{2} \mathring{Y}_{ki} \xi^l \xi_l =  \xi^l\varepsilon_{kl} \Gamma_i +\frac{i}{2} \mathring{Y}_{li} \xi_k \xi^l\;. 
\end{align}
Further, using \eqref{Yij}, this equation \eqref{YGGamma} can be rewritten as
\begin{align} \label{Y}
     &\frac{i}{2} \mathring{Y}_{ki} \xi^l \xi_l =  \xi^l\varepsilon_{kl} \Gamma_i + \xi_k \varepsilon_{ij} \Gamma^j\;.
\end{align}
The above equation \eqref{Y} can be split into two independent components; one symmetric in $i$ and $k$ and the other antisymmetric in $i$ and $k$. The symmetric part allows us to solve for $\mathring{Y}_{ik}$ as follows
    \begin{align} 
    \mathring{Y}_{ik}&= - \frac{2i}{\xi^m \xi_m} \bigg( \Gamma_{(i} \varepsilon_{k)l} \xi^l+ \xi_{(k} \varepsilon_{i)j}\Gamma^j \bigg) \;.
    \end{align}
Using the definition of $\Gamma_i$ in \eqref{Gammadef}, we can obtain the explicit composite expression for $\mathring{Y}_{ij}$ as 
\begin{align}\label{Yexplicit_sol}
   \mathring{Y}_{ik} &= \frac{2i}{\xi^n \xi_n}  \xi^l\varepsilon_{l(k} \bigg( \square_c \xi_{i)} - \frac{i}{2} \varepsilon_{i)j} \mathring{\bar{\Omega}}^j \theta_R - \frac{i}{2} \mathring{\bar{\Omega}}_{i)} \psi_L +\frac{1}{2} \bar{\chi}_{i)} \theta_R  - \frac{1}{2} \varepsilon_{i)j} \bar{\chi}^j \psi_L + \frac{1}{2}D \xi_{i)}  -  \lvert X \rvert ^2 \xi_{i)} \bigg) \nonumber \\ & +  \frac{2i}{\xi^m \xi_m} \varepsilon_{n(i} \xi_{k)} 
   \bigg( \square_c \xi^n + \frac{i}{2} \varepsilon^{nj} \mathring{\bar{\Omega}}_j \theta_L + \frac{i}{2} \mathring{\bar{\Omega}}^n \psi_R +\frac{1}{2} \bar{\chi}^n \theta_L  - \frac{1}{2} \varepsilon^{nj} \bar{\chi}_j \psi_R + \frac{1}{2}D \xi^n -  \lvert X \rvert ^2 \xi^n \bigg) \;.
\end{align}\\
The antisymmetric part of \eqref{Y}, can be written as,   
 \begin{align}\label{xi_anti_symm}
D^a \bigg( \xi^i D_a \xi_i - \xi_i D_a \xi^i + \frac{1}{2} \bar{\theta}_R \gamma_a \theta_L + \frac{1}{2} \bar{\psi}_L \gamma_a \psi_R \bigg) = 0 \;.
 \end{align}
We can take the dual of the 1-form inside the covariant derivative in \eqref{xi_anti_symm} and interpret \eqref{xi_anti_symm} as a Bianchi identity for the 3-form field strength of a two form gauge field $B_{\mu \nu}$. Precisely, equation \eqref{xi_anti_symm} can be re-written as:
\begin{align} \label{bianchi}
D_{[a}H_{bcd]} =0\;,  
\end{align}
where we have defined, 
\begin{align} \label{H}
    \frac{1}{3!} \varepsilon_{abcd}H^{bcd}& = \xi^i D_a \xi_i - \xi_i D_a \xi^i + \frac{1}{2} \bar{\theta}_R \gamma_a \theta_L + \frac{1}{2} \bar{\psi}_L \gamma_a \psi_R\;.
\end{align}
Here, we have used the Levi-Civita $\varepsilon_{abcd}\equiv i\epsilon_{abcd}$, where $\epsilon_{abcd}$ is the standard Levi-Civita with $\epsilon_{0123}=1$. By expanding the covariant derivatives on the R.H.S of \eqref{H}, we can interpret \eqref{H} as a constraint that determines the $U(1)$ gauge field $\mathring{W}_\mu$ of the central charge multiplet completely and makes it a composite gauge field as follows, 
\begin{align}\label{Wsol}
          \mathring{W}_a   & = \frac{i}{\xi^l \xi_l} \bigg( -\frac{1}{3!} \epsilon_{abcd}H^{bcd} + \xi^i \partial_a \xi_i - \xi_i \partial_a \xi^i + 2 \xi_i V_a^{i}{}_{j} \xi^{j} + \frac{1}{2} \bar{\theta}_R \gamma_a \theta_L + \frac{1}{2} \bar{\psi}_L \gamma_a \psi_R + \frac{1}{2} \xi^i \bar{\psi}_{ai} \theta_R \nonumber \\& - \frac{1}{2} \varepsilon_{ij} \xi^i \bar{\psi}^j_a \psi_L -  \frac{1}{2} \xi_i \bar{\psi}_{a}^i \theta_L + \frac{1}{2} \varepsilon^{ij} \xi_i \bar{\psi}_{aj} \psi_R \bigg)\;.
\end{align}
 
Using \eqref{H}, we can obtain the $Q$ and $S$- supersymmetry transformations of the 3-form field strength $H_{abc}$ as shown below:
\begin{align}\label{Htransf}
    \delta_Q \bigg( \frac{1}{3!} \varepsilon_{abcd}H^{bcd} \bigg)& = \bar{\epsilon}_i \gamma_{ab} D^b ( \theta_R \xi^i) + \varepsilon_{ij} \bar{\epsilon}^i \gamma_{ab} D^b (\psi_L \xi^j) +\frac{1}{16} \varepsilon_{ij} \xi^i \bar{\theta}_R \gamma \cdot T^{+} \gamma_a \epsilon^j\nonumber \\
    & +\frac{1}{16} \xi^i \bar{\psi}_L \gamma \cdot T^- \gamma_a \epsilon_i  - \text{h.c}\;, \nonumber \\
     \delta_S \bigg( \frac{1}{3!} \varepsilon_{abcd}H^{bcd} \bigg)& = - \frac{3}{2} \xi^i \bar{\theta}_R \gamma_a \eta_i + \frac{3}{2}\varepsilon_{jk} \xi^j \bar{\psi}_L \gamma_a \eta^k - \text{h.c}\;.
\end{align} 
One can use the above transformation rules \eqref{Htransf} and the Bianchi identity \eqref{bianchi} to obtain the supersymmetry transformations of the gauge field $B_{\mu \nu}$. This gauge field comes with its own vector gauge transformation parametrized by $\lambda_\mu$. The complete gauge transformations of $B_{\mu\nu}$ are as follows:
\begin{align}\label{Btrans}
    \delta B_{\mu \nu} & = \bar{\epsilon}_i \gamma_{\mu \nu} \theta_R \xi^i - \varepsilon_{ij} \bar{\epsilon}^i \gamma_{\mu \nu} \psi_L \xi^j -4 \xi_i \xi^j \bar{\epsilon}_j \gamma_{[\mu} \psi_{\nu]}^i + 2 \xi^i \xi_i \bar{\epsilon}_j \gamma_{[\mu} \psi_{\nu]}^j + \text{h.c} \nonumber \\
    & + 2 \partial_{[\mu} \lambda_{\nu]} \;.
\end{align}
From the above transformations of $B_{\mu \nu}$, the explicit form of $H_{\mu\nu\rho}$ can be obtained.
\begin{align}\label{HB}
     H_{\mu \nu \rho}= 3 \partial_{[\mu} B_{\nu \rho]}+ &\left( -  \frac{3}{2} \xi^i \bar{\psi}_{[\mu i} \gamma_{\nu \rho]} \theta_R  + \frac{3}{2} \xi^j\varepsilon_{ij} \bar{\psi}^i_{[\mu} \gamma_{\nu \rho ]} \psi_L+ 6 \xi_i \xi^j \bar{\psi}_{[\mu j} \gamma_{\nu} \psi_{\rho]}^i\right.\nonumber \\
     &\left.- 3 \xi^i \xi_i \bar{\psi}_{[\mu j} \gamma_{\nu} \psi_{\rho]}^j + \text{h.c}\right) \;.
\end{align}
One can check the results \eqref{Btrans} by using \eqref{HB} and obtaining the supersymmetry transformations \eqref{Htransf} and Bianchi-identity \eqref{bianchi}, where one needs to use the unconventional curvature constraints of \cite{Butter:2017pbp}.

Let us summarize our results. The field equations of the hypermultiplet \eqref{eom n=2 hyper} have been used to solve for the gauge field $\mathring{W}_{\mu}$, gaugino $\mathring{\Omega}_i$ and the auxiliary field $\mathring{Y}_{ij}$ of the central charge multiplet which becomes composite and are determined completely in terms of the hypermultiplet fields and a newly introduced 2-form gauge field $B_{\mu\nu}$. The complex scalar $X$ of the central charge multiplet remains independent and this together with the hypermultiplet fields as well as the 2-form gauge field $B_{\mu\nu}$ form an $8+8$ off-shell multiplet which we refer to as the Scalar-Tensor multiplet because of the presence of scalar fields and a 2-form tensor field in the multiplet. 

The counting of the off-shell degrees of freedom for this multiplet goes as follows. There are two Majorana spinors ($\psi$ and $\theta$) each carrying $4$ off-shell degrees of freedom. Hence the total number of off-shell fermionic degrees of freedom are $8$. There is an $SU(2)$ doublet of complex scalars ($\xi^i$) carrying $4$ real components. There is an $SU(2)$ singlet of complex scalar $X$ carrying $2$ real components. There is a anti-symmetric tensor gauge field $B_{\mu\nu}$ carrying $6$ real components. The bosonic gauge symmetries realized on this multiplet are the vector gauge transformations parametrized by $\lambda_\mu$ which are $3$ in number because of the identification $\lambda_\mu\sim \lambda_\mu+\partial_\mu\tilde{\lambda}$. We also have the $U(1)$ central charge transformation that is realized on this multiplet with a composite gauge field $\mathring{W}_\mu$. This gauge symmetry takes away $1$ degree of freedom. Finally subtracting the $4$ gauge symmetries from the $12$ bosonic components, we get $8$ bosonic off-shell degrees of freedom. 

The details of this mutliplet are presented in the next subsection.
 \subsection{Field content and supersymmetry transformations of the $\mathcal{N}$ = 2 Scalar-Tensor multiplet} \label{scalar tensor}

 The field contents of the multiplet together with their various properties are given in the table below:

 \begin{table}[h!]
	\centering
	\centering
	\begin{tabular}{ |p{1cm}|p{2cm}|p{1.5cm}|p{1.5cm}|p{2cm}|p{1.5cm}| }
		\hline
		Field & Type& $SU(2)_R$ irrep &$w$ (Weyl weight)&$c_A$ (chiral or $U(1)_R$ weight)& $c_z$ (central charge)\\
		\hline
		$\psi_R$& Fermion &\bf{1}&$3/2$&$-1/2$ & $1/2$\\
        $\theta_L$& Fermion & \bf{1} &$3/2$&$1/2$ & $1/2$\\	$\xi_i$&Boson&${\bf{2}}$&$1$&$0$&$-1/2$\\
		$B_{\mu\nu}$&Boson (Tensor Gauge field)&\bf{1}&$0$&$0$&$0$\\
  $X$ & Boson & \bf{1}& $1$& $-1$&$0$\\
		\hline
	\end{tabular}
	\caption{Field contents of the $\NN=2$ scalar-tensor multiplet}
	\label{scalar-tensor}	
\end{table}

 
 The $Q$ as well as $S$- supersymmetry transformations of this multiplet are given below, 
\begin{align}
\delta \psi_{R}&= - 2 \varepsilon_{jk} \cancel{D} \xi^k \epsilon^j - 2i X \xi^k \epsilon_k + 2 \varepsilon_{jk} \xi^j \eta^k \;,
\nonumber \\
\delta \theta_L &= - 2 \cancel{D} \xi^i \epsilon_i  - 2i \varepsilon_{jk} \bar{X} \xi^j \epsilon^k  - 2 \xi^i \eta_i \;,
\nonumber \\
\delta \xi_i&=-\bar{\epsilon_i} \theta_R + \varepsilon_{ij} \bar{\epsilon}^j \psi_L \;, \nonumber \\
\delta B_{\mu \nu} & = \bar{\epsilon}_i \gamma_{\mu \nu} \theta_R \xi^i - \varepsilon_{ij} \bar{\epsilon}^i \gamma_{\mu \nu} \psi_L \xi^j -4 \xi_i \xi^j \bar{\epsilon}_j \gamma_{[\mu} \psi_{\nu]}^i + 2 \xi^i \xi_i \bar{\epsilon}_j \gamma_{[\mu} \psi_{\nu]}^i + \text{h.c} \;, \nonumber \\
\delta X & =\bar{\epsilon}^i \mathring{\Omega}_i \;.
\end{align} 
The only composite field that appear explicitly in the above transformations is $\mathring{\Omega}_i$ given in \eqref{omegasol}. The composite gauge field $\mathring{W}_{\mu}$ given in \eqref{Wsol} will only appear inside the covariant derivatives of $\xi^i$. The composite gauge field $\mathring{Y}_{ij}$ given in \eqref{Yexplicit_sol} never appears in the above transformations but can appear in the transformations of other composite fields such as $\mathring{\Omega}_i$ which might be crucial to check the off-shell closure of the supersymmetry algebra on this multiplet.

 \section{Conclusion and Future Directions}\label{conc}
In this paper we discussed the construction of a new multiplet in $\NN=2$ conformal supergravity that contain scalar fields and a tensor gauge field. We referred to this multiplet as a scalar-tensor multiplet. This is obtained by a series of procedures highlighted as follows. Firstly, we employed the method of supersymmetric truncation on $\mathcal{N}=3$ conformal supergravity multiplets to obtain $\mathcal{N}=2$ conformal supergravity multiplets. The $\NN=3$ Weyl multiplet reduces to already known $\NN=2$ multiplets such as the $\mathcal{N}=2$ Weyl multiplet (\ref{dict weyl-weyl}) and the $\mathcal{N}=2$  off-shell vector multiplet (central charge multiplet) (\ref{dict weyl-vector}). Truncation of the on-shell $\mathcal{N}=3$ vector multiplet gave the $\mathcal{N}=2$ on-shell vector multiplet (\ref{vector to vector}) and an on-shell scalar multiplet (\ref{vector-hyper}), which was identified with the $\NN=2$ hypermultiplet with a broken rigid $SU(2)$ and a non trivial central charge transformation. The non trivial central charge transformation of the hypermultiplet allowed us to interpret its field equations as constraints on the central charge multiplet fields, which turned some of the fields of the central charge multiplet composite. The remaining fields from the central charge multiplet together with the hypermultiplet fields and a 2-form tensor gauge field constituted a new $8+8$ components off-shell multiplet of $\mathcal{N}=2$ conformal supergravity (\ref{scalar tensor}), which we named as the scalar-tensor multiplet.

Our work highlights the effectiveness of supersymmetric truncation in revealing new multiplets in conformal supergravity. As mentioned before, in four dimensions, $\mathcal{N}=4$ to $\mathcal{N}=3$ truncation and $\mathcal{N}=2$ to $\mathcal{N}=1$ truncation are already discussed in literature. These, together with the $\mathcal{N}=3$ to $\mathcal{N}=2$ truncation given in section \ref{truncation} completes this chain of supersymmetric truncations. Additionally it allowed us to get a new multiplet in $\NN=2$ conformal supergravity.

In \cite{Theis:2003jj}, the authors have discussed about couplings between scalar fields and tensor gauge fields by considering a multiplet which they also refer to as the scalar-tensor multiplet. However, their scalar tensor multiplet has been obtained by dualizing some of the scalar fields in an on-shell hypermultiplet and hence it is an on-shell multiplet. The scalar tensor multiplet obtained in this paper is an off-shell combination of an off-shell vector multiplet (the central-charge multiplet) and an on-shell hypermultiplet. This multiplet has a non-trivial central charge transformation with a dependent gauge field $\mathring{W}_{\mu}$. Hence, it is different from the scalar-tensor multiplet presented in \cite{Theis:2003jj}, although they share the same name.  

There are other multiplets known in the literature such as the linear multiplet \cite{deWit:1980lyi} which also has an anti-symmetric tensor gauge field along with scalar fields and a doublet of Majorana spinors carrying $8+8$ off-shell degrees of freedom. However, the off-shell structure of the linear multiplet and the scalar tensor multiplet derived in this paper are completely different as mentioned below. 
\begin{enumerate}
    \item 
The only gauge symmetries, apart from the gauge symmetries associated with the superconformal algebra, realized on the linear multiplet are the vector gauge transformations associated with the tensor gauge field. Whereas in the scalar tensor multiplet, there is an extra $U(1)$ central charge transformation that acts non-trivially on the multiplet and the corresponding gauge field $\mathring{W}_\mu$ is composite. 
\item 
In the linear multiplet, the fermionic degrees of freedom are encoded in an $SU(2)$ doublet of Majorana spinors, whereas in the scalar tensor multiplet, they are encoded in two $SU(2)$ singlets of Majorana spinors. The Weyl weight of the fermions in the linear multiplet is $5/2$ and the chiral Weight of the left-chiral components of the fermions in the linear multiplet is $1/2$ \cite{deWit:1980lyi}. As can be seen in Table-\eqref{scalar-tensor}, the Weyl weights of the fermions in the scalar-tensor multiplet are different from that of the fermions in the linear multiplet.
\item
In the linear multiplet, the bosonic degrees of freedom are encoded in an $SU(2)$ triplet of pseudo-real scalar fields $L_{ij}$, an $SU(2)$ singlet of complex scalar field $G$ along with the tensor gauge field $B_{\mu\nu}$. The Weyl weights of these fields are $2$, $3$ and $0$ respectively and the chiral weights are $0$, $1$ and $0$ respectively \cite{deWit:1980lyi}. As can be checked from Table-\eqref{scalar-tensor}, the $SU(2)$ irreps as well as the Weyl and chiral weights are different for some of the bosonic field contents of the scalar-tensor multiplet.
\end{enumerate}
Because of the above-mentioned differences in the off-shell structure, the linear multiplet cannot be used as a single compensating multiplet to obtain $\NN=2$ Poincar{\'e} supergravity from conformal supergravity. It is always combined with the $\NN=2$ vector multiplet to compensate for the extra symmetries present in $\NN=2$ conformal supergravity in order to obtain $\NN=2$ Poincar{\'e} supergravity \cite{deWit:1982na}. However, the scalar-tensor multiplet that we have discussed in this paper is an off-shell combination of a central charge multiplet and an on-shell hypermultiplet. Hence, unlike the linear multiplet, this multiplet looks like a potential candidate to be used as a single compensating multiplet for getting $\NN=2$ Poincar{\'e} supergravity from conformal supergravity. It has all the fields necessary for compensating the extra symmetries of conformal supergravity. The field $\xi^i$ (which has 4 degrees of freedom) can be used to completely fix the $SU(2)$ R-symmetry and leave behind a real scalar degree of freedom. The field $X$ can also be used to compensate for the $U(1)$ R-symmetry as well as the dilatation symmetry. The Majorana spinors $\psi$ and $\theta$ will together act as a compensator for $S$-supersymmetry. However, in the scalar tensor multiplet the graviphoton $W_{\mu}$ is a composite field and it has been replaced by a tensor gauge field. Hence, the Poincar{\'e} supergravity that one may obtain using the scalar tensor multiplet as a compensator will not have an independent graviphoton. Instead, one will have a tensor gauge field. The construction of such a Poincar{\'e} supergravity using this scalar-tensor multiplet as a compensator and finding its relation with the $\NN=2$ Poincar{\'e} supergravity in the literature which has an independent graviphoton looks an interesting avenue for further research. The relation of such a construction with the principal version of $\NN=2$ Poincar{\'e} supergravity obtained in \cite{Galperin:1987ek,Galperin:2001seg} would be also interesting to see. We would like to address this in the future.

\acknowledgments
AA thanks Chennai Mathematical Institute for hospitality during the course of this work. We thank Soumya Adhikari, Abhinava Bhattacherjee, Madhu Mishra and Krishnanand K. Nair for useful discussions.

\appendix

\section{Conventions} \label{conventions}
Since the $\mathcal{N}=2$ multiplets in this paper are obtained via a supersymmetric truncation of $\mathcal{N}=3$ multiplets, they inherit the conventions used for the $\mathcal{N}=3$ multiplets (see the appendix of \cite{Hegde:2021rte}).
If we make the following changes to the $\mathcal{N}=2$ Weyl multiplet given in section \ref{N=2 weyl} we can reproduce $\mathcal{N}=2$ Weyl multiplet given in \cite{Butter:2017pbp} which uses unconventional constraints on the curvatures.

\begin{align}
    &V_{\mu}^i{}_j \longrightarrow -\frac{1}{2} V_\mu^i{}_j\;, \nonumber\\
    &T_{ab}^- \longrightarrow \frac{1}{2} \epsilon_{ij}T^{ij}_{ab}\;,\nonumber\\
&\delta_Q(\epsilon) \longrightarrow \delta_{Q}(\epsilon)+ \delta_K(\Lambda_{K \mu} =  \frac{1}{2} \epsilon^i \gamma_{\mu} \chi_i + h.c )\;,
\end{align}

where $\delta_K$ denotes a special conformal transformation. In order to obtain the $\NN=2$ Weyl multiplet in the conventional constraints \cite{Bergshoeff:1980is} one can do further field redefinitions as discussed in \cite{Butter:2017pbp}.

			
\bibliography{biblio}
\bibliographystyle{jhep}

\end{document}